\documentclass[twocolumn,prb,aps,superscriptaddress]{revtex4}
\usepackage{graphicx}
\usepackage{enumitem}
\usepackage{bm}

\newcommand{\TaS}{$2H$-TaS$_2$\,}
\newcommand{\NbSe}{$2H$-NbSe$_2$\,}
\newcommand{\TaSe}{$2H$-TaSe$_2$\,}
\newcommand{\Dcdw}{$\Delta_{\text{cdw}}\,$}
\newcommand{\Tcdw}{$\it{T}_{\text{cdw}}\,$}
\newcommand{\qcdw}{$\textbf{q}_{\text{cdw}}\,$}

\begin{document}

\title{Orbital selectivity causing anisotropy and particle-hole asymmetry in the charge density wave gap of \TaS
}
\author{J. Zhao}
\affiliation{Department of Physics, University of Virginia, Charlottesville, VA 22904, USA}
\author{K. Wijayaratne}
\affiliation{Department of Physics, University of Virginia, Charlottesville, VA 22904, USA}
\author{A. Butler}
\affiliation{Department of Physics, University of Virginia, Charlottesville, VA 22904, USA}
\author{V. Karlapati}
\affiliation{Department of Physics, University of Virginia, Charlottesville, VA 22904, USA}
\author{J. Yang}
\affiliation{Department of Physics, University of Virginia, Charlottesville, VA 22904, USA}
\author{C. D. Malliakas}
\affiliation{Department of Chemistry, Northwestern University, Evanston, IL 60208, USA}
\author{D. Y. Chung}
\affiliation{Materials Science Division, Argonne National Laboratory, Argonne, IL 60439, USA}
\author{D. Louca}
\affiliation{Department of Physics, University of Virginia, Charlottesville, VA 22904, USA}
\author{M. G. Kanatzidis}
\affiliation{Department of Chemistry, Northwestern University, Evanston, IL 60208, USA}
\affiliation{Materials Science Division, Argonne National Laboratory, Argonne, IL 60439, USA}
\author{J. van Wezel}
\affiliation{10H.H. Wills Physics Laboratory, University of Bristol, Tyndall Ave., Bristol BS8 1TL, UK}
\author{U. Chatterjee}\thanks{Correspondence to: Email: uc5j@virginia.edu (U.C.)}
\affiliation{Department of Physics, University of Virginia, Charlottesville, VA 22904, USA}
\date{\today}

\begin{abstract}
We report an in-depth Angle Resolved Photoemission Spectroscopy (ARPES) study on \TaS, a canonical  incommensurate Charge Density Wave (CDW) system. This study demonstrates that just as in related incommensurate CDW systems, \TaSe and \NbSe, the energy gap (\Dcdw) of \TaS is localized along the K-centered Fermi surface barrels and is particle-hole asymmetric.  The persistence of \Dcdw  even at temperatures higher than the CDW transition temperature \Tcdw  in \TaS, reflects the similar pseudogap (PG) behavior observed previously in \TaSe and \NbSe. However, in sharp contrast to \NbSe, where \Dcdw  is non-zero  only in the vicinity of a few ``hot spots'' on the inner K-centered Fermi surface barrels, \Dcdw  in \TaS is non-zero along the entirety of both K-centered Fermi surface barrels. Based on a tight-binding model, we attribute this dichotomy in the momentum dependence and the Fermi surface specificity of \Dcdw  between otherwise similar CDW compounds to the different orbital orientations of their electronic states that are involved  in CDW pairing. Our results suggest that the orbital selectivity plays a critical role in the description of incommensurate CDW materials. 

\end{abstract}
\maketitle

%main text-------------------------------------------------------------------------------------
%%-----------------------------------------------------------------A: Introduction-------------------------------------------------------------
\section{Introduction}
Layered transition metal dichalcogenides (TMDs) are highly sought-after materials for their extremely rich phase diagrams, which encompass diverse quantum states including metals, semiconductors, Mott insulators, superconductors and charge density waves (CDWs) \cite{Wilson1, Wilson2, Monceau, Friend_Yoffe, Kyle_Review}. \TaS, a prominent member of the TMD family, is an extremely versatile material by itself. In its pristine and bulk form, \TaS hosts an incommensurate CDW order with the wave-vector \qcdw$\sim (2/3-0.02)\,\Gamma$M \cite{BS4} and the \Tcdw $\sim$75K \cite{CDW_TaS2_1, CDW_TaS2_2}. The CDW order coexists with the superconductivity \cite{SC_TaS2_1, SC_TaS2_2, SC_TaS2_3} at temperatures lower than 0.8K. Like a number of  other TMDs \cite{UC_NATCOM, TaS2_PD1, TaS2_PD2, TaS2_GATE, TiSe2_Cu, TiSe2_Pd, TaSe2_Pd}, the CDW and superconducting properties of \TaS are also intertwined and can be tuned via various materials processing techniques such as chemical intercalation \cite{Cava_Cu_TaS2, Cu_TaS2_2,  Fe_TaS2, Ni_TaS2, Na_TaS2_1, Na_TaS2_2, Intercalation_TaS2_1, Intercalation_TaS2_2},  strain engineering \cite{pressure_TaS2} and exfoliation \cite{exfoliation_TaS2}. Furthermore, \TaS-based alloys, intercalated with 3$d$ elements \cite{3D_TaS2}, are shown to have great relevance in applications in magnetic devices --- for example, a very pronounced out-of-plane magnetocrystalline anisotropy emerges upon Fe intercalation \cite{PMA_TaS2}.

Studies of incommensurate CDW order in TMDs such as \NbSe and \TaS have attracted a lot of attention lately. For example, a series of spectroscopic measurements \cite{Borisenko_NbSe2, Kyle_NbSe2, JENI_NbSe2, ABHAY_NbSe2_1, ABHAY_NbSe2_2, CROMMIE_NbSe2, Feng_NbSe2, Feng_TaS2, Wang_OPTICS, OLSON_TaS2,  Mcelroy_NbSe2} have revealed that $\Delta_{\text{cdw}}$ of these compounds opens up only around specific regions of their underlying Fermi surfaces (FS's) \cite{Borisenko_NbSe2, Kyle_NbSe2}. Contrary to the traditional picture \cite{GRUNNER_BOOK}, FS nesting alone was shown not to be responsible for the CDW instability in \NbSe \cite{Borisenko_NbSe2, FRANK_NBSE2_PRL}. However, there are reports both for and against FS nesting alone as the driver of the CDW order in \TaSe \cite{Borisenko_TiSe2, NO_FS_NESTING_TaS2}. Moreover, $\Delta_{\text{cdw}}$ in \NbSe has been found particle-hole asymmetric \cite{UC_NATCOM, JENI_NbSe2, Flicker & JvW_1,Flicker & JvW_2}, and non-zero even at temperatures ($T$'s) greater than \Tcdw \cite{Borisenko_NbSe2, UC_NATCOM}, which resembles the enigmatic pseudogap (PG) behavior in underdoped cuprate high temperature superconductors (HTSCs) \cite{PG_BISCO1, PG_BISCO2, Timusk_PG}. The PG behavior has also been observed in \TaSe \cite{Borisenko_TiSe2}. Recently, a theoretical analysis of the CDW order in \NbSe showed that these intriguing observations can be modeled within a single theory based on strong electron-phonon (el-ph) coupling \cite{Flicker & JvW_1, Flicker & JvW_2}. Quite strikingly, the orbital character of the electronic states involved in the CDW formation as well as the momentum dependence of the el-ph coupling play equally significant roles in this model. 

In light of the above-described developments in our understanding of  incommensurate systems like \NbSe and \TaSe, and the close parallel between these systems and the cuprate HTSCs, an important question emerges: which of the above-described experimental observations are universal attributes of incommensurate CDW systems?  In order to address this, we present here a comprehensive study of the electronic structure of \TaS as a function of $T$ and momentum ($\mathbf{k}$). Unlike the cases of \NbSe and \TaSe, spectroscopic investigations on \TaS are rather limited. Angle Resolved Photoemission Spectroscopy (ARPES) studies on \TaS \cite{OLSON_TaS2} and a related material Na$_x$TaS$_2$  \cite{Feng_TaS2, Feng_NbSe2} have found non-zero \Dcdw only along selected regions of the underlying FS in each case. These were corroborated by the optics data \cite{Wang_OPTICS}.  

The objectives of this paper are:  (i) to interrogate whether \Dcdw is particle-hole symmetric or asymmetric;  (ii)  to examine the possible existence of  PG  behavior for $T>$\Tcdw; and (iii)  to investigate whether the experimental data bears any manifestation of orbital selective CDW pairing. We emphasize that these issues have not been explored in previous studies. 
Establishing the $\bf{k}$ and $T$ dependence  of \Dcdw and its FS specificity are vital to unveiling the mechanism of CDW order in \TaS. Moreover, a direct comparison of this information with that from \NbSe and \TaSe will be helpful to identify the universal traits of incommensurate CDW systems. 

Our ARPES data, combined with arguments based on a tight-binding model, establish that: (i) like in \NbSe, $\Delta_{\text{cdw}}$ in \TaS is particle-hole asymmetric and persists for $T>$\Tcdw ; (ii) in contrast to \NbSe, $\Delta_{\text{cdw}}$ in \TaS is clearly visible at each measured momentum location along both K-centered FS barrels; and (iii) the difference between the momentum anisotropy and the FS specificity of $\Delta_{\text{cdw}}$ in \TaS and that in \NbSe can be understood by comparing the orbital natures of the electronic states involved in CDW pairing.
 
%%----------------------------------------------------------B: Experimental Details---------------------------------------------------------
\section{Experimental Details}
We have conducted ARPES measurements using the 21.2 eV Helium-I line of a discharge lamp combined with a Scienta R3000 analyzer at the University of Virginia, as well as 75 eV and 22 eV synchrotron light equipped with a Scienta R4000 analyzer at the PGM beamline of the Synchrotron Radiation Center, WI. The angular resolution is $\sim$ 0.3 degree, and the total energy resolution is $\sim$ 8--15 meV. For $T$-dependent studies, data were collected in a cyclic way to ensure that there were no aging effects in the spectra. All experiments were performed in ultra high vacuum (better than $5\times10^{-11}$Torr, both in the helium lamp system and in the beamline). Single crystals of \NbSe and \TaS were grown using the standard iodine vapor transport method.  Conventional four-terminal configuration was employed for measuring the resistivity of \TaS single crystal sample in a Quantum Design Physical Properties Measurement System (PPMS). Electrical resistivity measurements indicate \Tcdw $\sim$ 75K (Fig. 1(g)), for \TaS, in agreement with previous studies \cite{SC_TaS2_1}.

%%-----------------------------------------------------------------C: Results:-------------------------------------------------------------
\section{Results}
\subsection{FS topology and nesting vectors} 
The first-principles calculations for both \TaS and \NbSe predict two closely spaced pairs of quasi-two-dimensional FS cylinders around the $\Gamma$ point as well as around the K point \cite{BS1, BS2, BS3, BS4}. These cylinders are double-walled due to the presence of two formula units per unit cell. Figs. 1(a) and 1(d) show the FS intensity maps of \TaS and \NbSe respectively, in their normal states. These FS intensity maps present the ARPES data at $\overline{\omega}=0$ as a function of the in-plane momentum components ${k_{x}}$ and ${k_{y}}$, where $\overline{\omega}$ is the electronic energy measured with respect to the chemical potential $\mu$. Notice that no symmetrization of ARPES data have been incorporated for constructing either of the FS intensity maps in Figs. 1(a) and 1(d). As expected \cite{BS1, BS2, BS3, BS4}, double-walled FS barrels around $\Gamma$ and K points can be observed in both compounds. The regions with high intensity along $\Gamma$-K, which are due to saddle bands, can also be noticed in both materials. However, the pancake like intensity profile around $\Gamma$ point, which is observed in \NbSe, is not detected in \TaS. All these observations are consistent with previous experiments on related compounds \cite{Borisenko_NbSe2, Feng_TaS2, OLSON_TaS2, Kyle_NbSe2, UC_NATCOM}.
\begin{figure}
\includegraphics[width=3.52in]{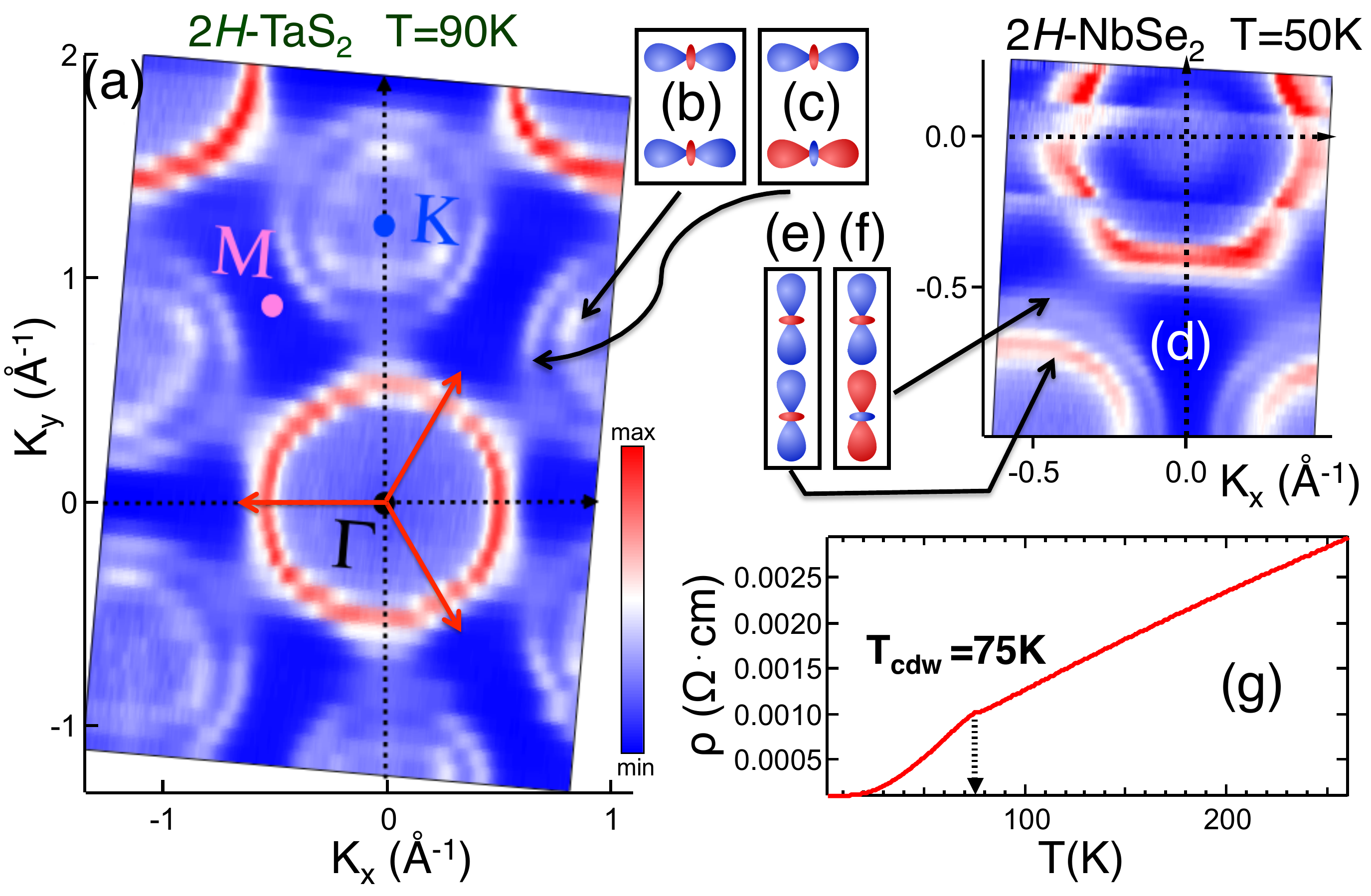}
\caption{(a) FS intensity map of a \TaS sample obtained using ARPES with photon energy $h\nu = 75\,$eV at $T=90\,$K. The red arrows correspond to the three primary CDW vectors $\mathbf{q_1}$, $\mathbf{q_2}$ and $\mathbf{q_3}$.  (b) and (c)  schematically indicate the dominant orbital contribution to the electronic states around the K point of \TaS, while (e) and (f) indicate that of \NbSe. (d) The FS  intensity map of a \NbSe sample at $T=50\,$K using $h\nu=22\,$eV. Note that the data has been integrated over an energy window of $10\,$meV to enhance the spectral features in both Figs. 1(a) and 1(d). (g) Resistivity ($\rho$) plotted against $T$ for  the \TaS sample. The CDW-induced anomaly, signaling \Tcdw$\sim 75\,$K, is identified by the discontinuous change in the slope of $\rho(T)$ curve, indicated by the black dashed arrow.}
% 1(h) Angle integrated photoemission data in normal emission from the \TaS sample, in which its valence bands as well as two of the Ta core level peaks are clearly visible.}
\end{figure}
In the case of a Peierls-like CDW instability \cite{GRUNNER_BOOK}, one expects the CDW wave-vector to span nearly parallel regions of the FS. Fig. 1(a) shows that although the FS of \TaS has a number of nearly parallel regions, their separations do not agree with the magnitude of \qcdw. For instance, both FS barrels around the $\Gamma$ point are too large in size for being self-nested by any of the three primary CDW wave vectors $\mathbf{q_1}$, $\mathbf{q_2}$ and $\mathbf{q_3}$ (shown by red arrows in Fig. 1(a)). As in the case of \NbSe \cite{Borisenko_NbSe2}, a simple FS nesting is therefore expected not to play a key role in the CDW formation in \TaS.

\subsection{CDW energy gap} 
In order to interrogate the momentum structure of \Dcdw along the K-centric FS barrels of \TaS, we focus on ARPES data as a function of $\overline{\omega}$ at specific momentum values, known as energy distribution curves (EDCs), at $T=45$K $<$ \Tcdw (Figs. 2(a), 2(b) and 2(c)). 
%In this context, the $\Gamma$-centric FS barrels were reported to be gapless in the CDW state \cite{Feng_TaS2, OLSON_TaS2}. 
The momentum locations of the EDCs in Figs. 2(a), 2(b) are marked in the FS intensity map around the K point in Fig. 2(d). To determine the presence of CDW energy gap, the effects of the Fermi function (FF) and energy resolution are to be eliminated from the EDCs. This can be accomplished to a good approximation after the EDCs are divided by the resolution broadened Fermi function \cite{JC_ARPES_REVIEW, ZX_ARPES_REVIEW}. We refer the method to take into account of fermi cut-off and energy resolution via division of EDCs by resolution broadened FF as Method 1.

The $\overline{\omega}$ location of the quasiparticle peak in the FF divided EDC can be seen to be below zero at each measured momentum point along the K-centered FS barrels (Figs. 2(e), 2(f)). This means that both K-centered FS barrels are gapped. To visualize this better, we stacked EDCs before and after division by resolution broadened FF at equal momentum spacing in Figs. 2(c) and 2(g) respectively along the momentum cut, which crosses through both barrels and is denoted by the black arrow in Fig. 2(d). It is apparent that the location of the quasiparticle peak of each FF divided EDC along the marked cut in Fig. 2(g) is below zero, which in turn is an evidence for the presence of an energy gap along  both FS barrels. It is further noted that the minima of the FF divided EDCs in Figs. 2(e), 2(f) and 2(g) are away from $\overline{\omega}=0$, which is a manifestation of the fact that  \Dcdw along the K-centered FS barrels is particle-hole asymmetric. A similar particle-hole asymmetry has been observed in \NbSe \cite{UC_NATCOM, JENI_NbSe2}. This can be contrasted with a superconducting energy gap, in which emergent particle-hole symmetry ensures the spectral minimum to be at $\overline{\omega}=0$.

In recent ARPES works \cite{Johnson}, the effects of energy and momentum resolution were analyzed by adopting the Lucy-Richardson iterative technique, which is different from the division of an EDCs by the resolution broadened FF \cite{JDR_THESIS, JDR_PAPER}. We refer this as Method 2 in Fig. 2. For the purpose of comparison, we plot the EDCs obtained after adopting Method 1 and Method 2 in Fig. 2(h). The momentum location of this particular EDC is denoted by the red triangle on the outer K-centric FS barrel in Fig. 2(d). Irrespective of the method we apply, Fig. 2(d) alludes particle-hole asymmetric \Dcdw in \TaS.

\begin{figure}[h]
\includegraphics[width=3.5in]{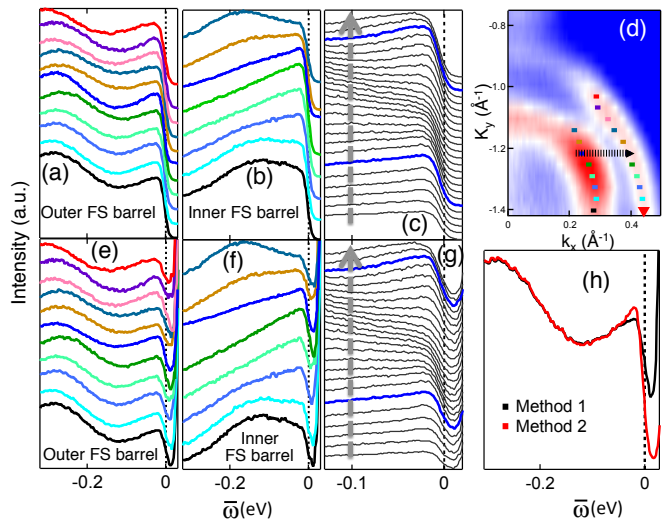}
\caption{(a and b) Raw EDCs at the momentum locations marked on the FS barrels around the K point in (d). (c) Raw EDCs at equal  momentum \textbf{\emph{k}} spacing along the marked cut (shown by a dashed arrow) in (d). (d) FS intensity map of a \TaS sample at T=45 K (h$\nu$=22 eV). (e), (f) and (g) are same as (a), (b) and (c), but after division by the resolution broadened FFs. (h) Comparison of the EDCs at the momentum location pointed by the red triangle in (d) after employing Method 1 and Method 2. Note that the markers of the momentum locations in (d) are color coded to be in conformity with the corresponding EDCs in (a), (b), (e), and (f). The black dotted lines denote $\overline{\omega}=0$.
}
\end{figure}

The non-zero value of \Dcdw along the entire inner and outer  K-centered FS barrels of \TaS is qualitatively different from the momentum dependence of \Dcdw in \NbSe, where \Dcdw is reported to be non-zero only in the neighborhood of the specific hot spots on the inner FS barrel \cite{Borisenko_NbSe2, Kyle_NbSe2}. We describe in section D how this can be understood in terms of the difference in the orbital character of electronic states in these two compounds.

%\subsection{Signature of backbending in electronic dispersion}
\begin{figure}[h]
\includegraphics[width=3.3in]{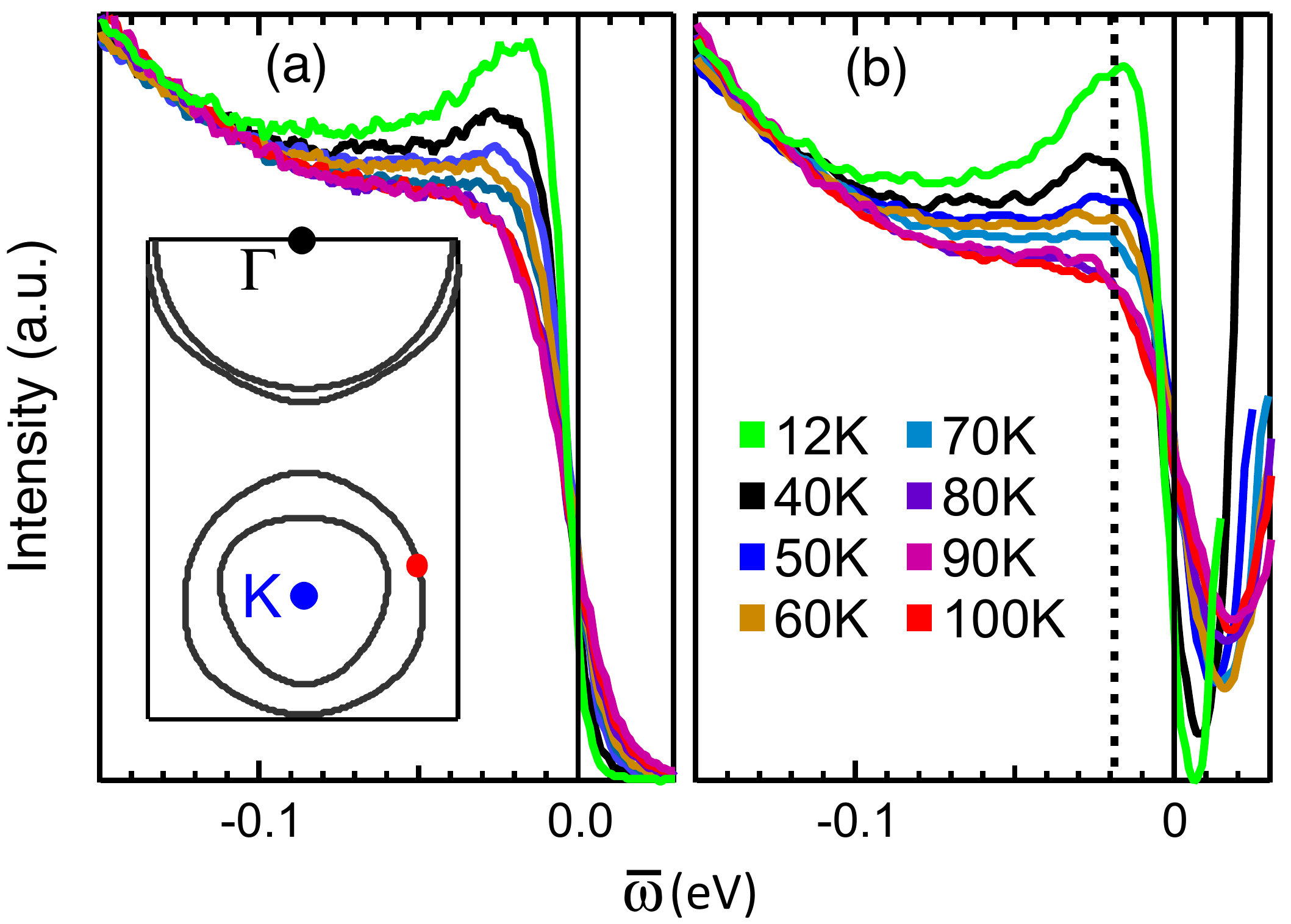}
\caption{(a) EDCs as a function of temperature going through \Tcdw. (b) EDCs in (a) after division by resolution broadened FF. The inset in (a) shows the schematic FS of \TaS, and the red dot on it marks the momentum location of these EDCs.}
\end{figure}

%The CDW order in a physical system leads to particle-hole mixing, which alters the electronic dispersion, i.e., the relation between electronic energy and momentum, in the CDW state compared to the one in the state without CDW order [\ref{GRUNNER_BOOK}]. In a CDW state with ordering wave vector $\mathbf{q}$, electronic states with momentum $\mathbf{k}$ are coupled to those at momentum $\mathbf{k+q}$. Consequently, the electronic dispersion is modified, and becomes $E_{k} = \frac{1}{2}(\epsilon_k+\epsilon_{k+q}) \pm \sqrt{\frac{1}{4}(\epsilon_k - \epsilon_{k+q})^2 + \Delta_k^2}$, where $\Delta_k$ is the CDW energy gap and $\epsilon_k$  is the normal state dispersion. The dispersion of the low-energy state in the CDW state thus starts to deviate from $\epsilon_k$ as it approaches $\epsilon_{k+q}$, reaches a maximum at the $\mathbf{k}$ value for which $\epsilon_k=\epsilon_{k+q}$, and then bends back to follow $\epsilon_{k+q}$ instead of $\epsilon_k$ (see Fig. 3a for a sketch in the case of a simple one-dimensional CDW system with a Peirls instability). Such backbending in the electronic dispersion, commonly known as the Bogoliubov dispersion [\ref{TINKHAM_SC}], also happens in a superconductor for T$<\text{T}_{\text{c}}$. The Bogoliubov dispersion has directly been observed in previous ARPES studies of cuprate HTSCs [\ref{JC_BB, MATSUI_BB, AK_BB}] and \NbSe [\ref{UC_NATCOM}]. Following the peak positions in the EDCs of \TaS in Fig. 3b show that it also has the expected back-bending behavior for T$<$\Tcdw.

\subsection{$T$-dependence of CDW gap and coherence}
The $T$-dependent ARPES data from \TaS is examined in Figs. 3(a) and 3(b). In Fig. 3(a), we show the EDCs at the momentum location indicated by a red dot in the inset of Fig. 3(a), while Fig. 3(b) displays those after division by resolution broadened FF's.  With increasing temperature, the intensity of the coherence peak in the FF divided EDCs in Fig. 3(b) is diminished, but  its $\overline{\omega}$ location remains approximately constant. Above \Tcdw, the coherence peak disappears and the spectra stop evolving with $T$. Although the spectra for T$>$\Tcdw do not have well-defined peaks, they do have a clearly discernible ``kink'' feature, defined as a discontinuous change in slope. From Fig. 3(b), it is apparent that the $\overline{\omega}$ location of this kink in the spectra at $T>$\Tcdw is approximately constant, and it is same as that of the coherence peaks for $T<$\Tcdw. The decrease in intensity of the coherence peak with increasing temperatures can be visualized from Fig. 3(a) as well.

We cannot determine the exact magnitude of the CDW energy gap from our ARPES measurements because of the particle-hole asymmetry of $\Delta_{\text{cdw}}$. Nevertheless, the fact that the peak/kink structures in the FF divided ARPES spectra are positioned at energy values $\overline{\omega}<0$, evidences that a non-zero $\Delta_{\text{cdw}}$ persists even for temperatures $T>$\Tcdw. The energy gap remains particle-hole asymmetric for all measured temperatures. Moreover, there is a loss of single-particle coherence  at \Tcdw, indicated by the disappearance of a peak from the spectra for $T>$\Tcdw. Similar behavior was observed in \NbSe \cite{UC_NATCOM} and underdoped Bi2212 HTSC \cite{UC_PNAS_PD} as $T$ is increased through \Tcdw and $\text{T}_{\text{c}}$ respectively. These observations suggest that the disappearance of the CDW order at \Tcdw in \TaS occurs due to loss of long range phase coherence, as suggested for \NbSe also \cite{UC_NATCOM}. Such an interpretation agrees with the fact that transmission electron microscopy measurements on \TaS show the presence of short range CDW order at temperatures above \Tcdw \cite{Cava_Cu_TaS2}. In this scenario, the transition to the PG state in both incommensurate CDW systems and underdoped cuprate HTSCs can thus be viewed as a transition from a coherent and gapped electronic state to an incoherent and gapped one.

\subsection{Dichotomy between \NbSe and \TaS}
Whereas the K-centered FS barrels of  \NbSe consist primarily of $d_{z^2}$ orbitals aligned along the crystallographic $c$-axis, as shown schematically in Figs. 1(e) and 1(f), the smaller size of the pockets in \TaS, and the corresponding proximity to a high-symmetry point, cause their electronic states to approach $d_{z^2}$ orbitals rotated onto the crystallographic $ab$-plane \cite{Whangbo & Canadell, BS4, Flicker & JvW_1, JvW}. The in-plane orbital configuration (shown schematically in Figs.1(b) and 1(c)) has been used before to argue that \TaS contains hidden one-dimensional order \cite{Whangbo & Canadell}, as well as a specific type of orbital order \cite{JvW}.

\begin{figure}[t]
\includegraphics[width=\columnwidth]{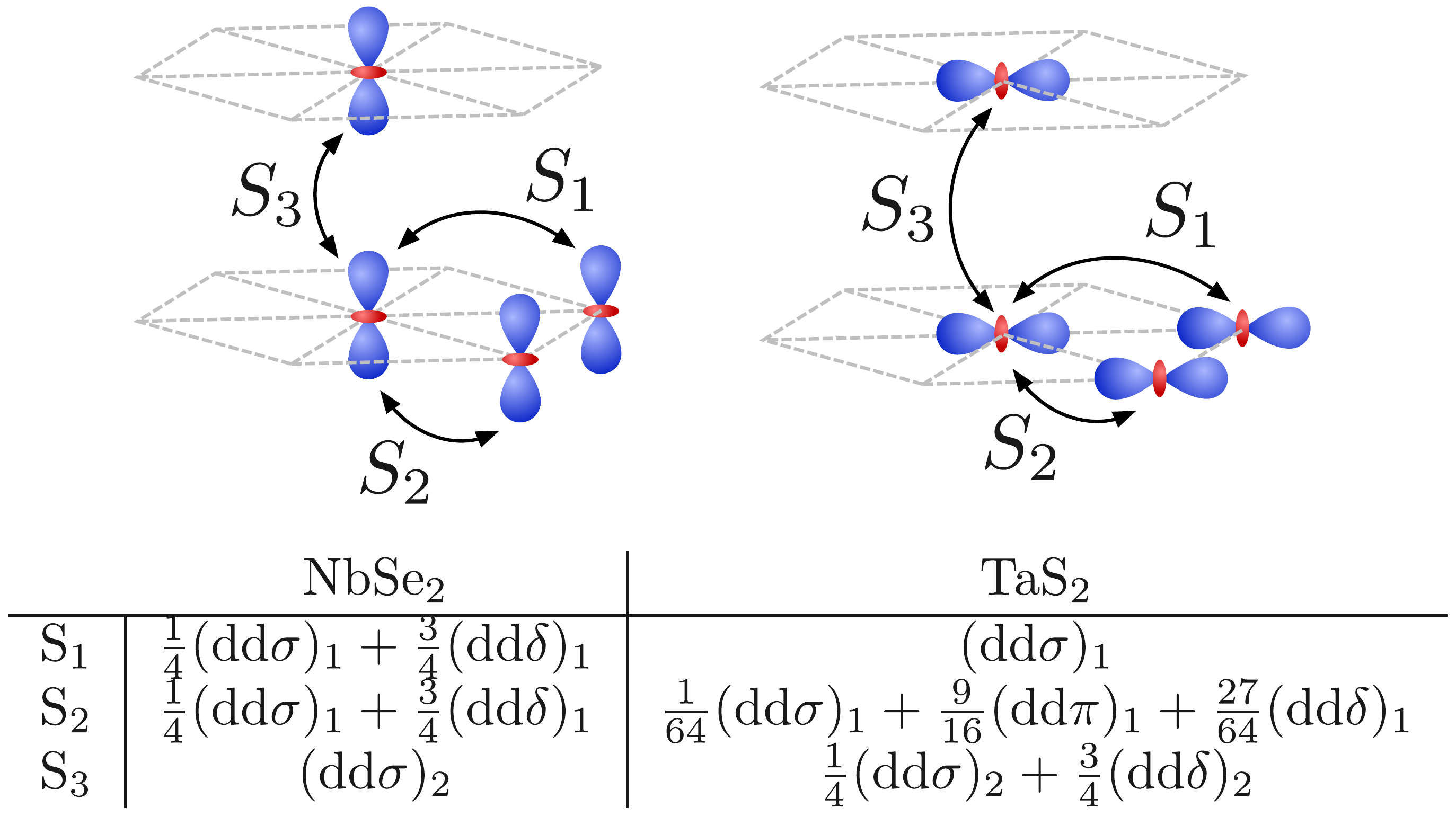}
\caption{Overlap integrals relevant to a tight-binding description of the electronic band structure. The different orbital orientations in the K-centered FS barrels of \NbSe and \TaS are shown schematically in the top row. They result in different entries of the overlap matrix $S(k)$, which (at the K-point, and taking into account nearest neighbor overlaps only) has the value $1+2S_2-S1$ for its diagonal elements, and $S_3$ for its off-diagonal elements. The squared ratio of these elements determines the ratio of CDW gap sizes in the concentric barrels.}
\label{overlaps}
\end{figure}

The orbital character of the electronic states will influence the strengths of the el-ph coupling in each K- centered FS barrel. In order to estimate the size of this effect, a tight-binding fit to the electronic band structure is required \cite{Varma 1979}, which involves the same set of overlap integrals between neighboring $d$-orbitals in the case of \TaS as it does for \NbSe. The relation between the two cases is shown in Table~\ref{overlaps}. The ratio between diagonal (in-plane) and off-diagonal (out-of-plane) elements in the overlap matrix $S(k)$ determines the relative strength of the el-ph coupling matrix elements in the two K-centered FS barrels, and hence the ratio of CDW gap sizes~\cite{Flicker & JvW_1, Flicker & JvW_2}. 

Taking all overlap integrals to be zero, except $(\text{dd}\sigma)_1=(\text{dd}\sigma)_2\approx 0.5$, our model yields a gap ratio between inner and outer barrels in \NbSe of $6.8$, while in \TaS it is much smaller, $\sim$ $2.7$. The contrast in gap ratios that emerges already at this simplest possible level of approximation, is in accord with the observed dichotomy in CDW gap structure between \TaS and \NbSe. The ARPES data in \TaS clearly exhibits a CDW gap along both FS barrels around the K point, consistent with a small ratio of gap sizes. On the contrary, the larger ratio for \NbSe results in ARPES data \cite{Borisenko_NbSe2, Kyle_NbSe2} finding \Dcdw to be restricted to the inner K-centered FS barrels only.

In addition to explaining the apparent difference between the momentum profiles of the CDW gap in \NbSe and \TaS, this model based on the presence of strong electron-phonon coupling also provides a natural explanation for the existence of a pseudogap phase at temperatures above \Tcdw. Analogous to the difference between weak coupling (BCS) and strong coupling (Eliashberg) theories for superconductivity, a strong-coupling CDW phase generically melts its charge order through increased phase fluctuations, rather than a suppressed CDW amplitude \cite{Flicker & JvW_1, UC_NATCOM}. Since the CDW gap is directly proportional to the local amplitude of the order parameter, it is present both in the short range fluctuating phase at $T$$>$\Tcdw and in the phase with long range CDW  order at $T$$<$\Tcdw.

\section{Conclusions}
Among the family of the TMDs, \TaS and \NbSe are both considered prototypical incommensurate CDW materials, where the experimental signatures of the CDW order are similar in various ways. Both have particle-hole asymmetric gaps on only some of their FS sheets, and in both cases the \Dcdw persists into a pseudogap phase above \Tcdw. Furthermore, FS nesting alone doesn't seem to be the driver of the CDW instability in both compounds. On the other hand, we also have established some pronounced differences. \Dcdw in \NbSe  is non-zero only near a few hot spots within a single K-centric FS barrel, while it is  non-zero along both barrels of \TaS. This dichotomy between \NbSe and \TaS can be realized in terms of the difference in the orbital structures of their electronic states in the vicinity of their Fermi levels. The different orientation of the \TaS states as compared to those of \NbSe directly affects the relative size of the el-ph coupling on the concentric FS barrels. Within a strong-coupling description of the CDW formation, the result is a strongly barrel-dependent gap size in \NbSe, and an approximately uniform gap in \TaS, in agreement with our current ARPES observations and previously published data. Additionally, the description in terms of a strong rather than weak coupling scenario implies that the location of the CDW gap depends on the momentum variations of the electron-phonon coupling rather than just the electronic structure, and hence will generically be centered slightly away from the FS, making it particle-hole asymmetric. Additionally, the order in strong-coupling theories is destroyed at \Tcdw by phase fluctuations, leaving the local gap size non-zero, and hence giving rise to a pseudogap phase above \Tcdw. Given the striking agreements between the results of the strong coupling approach and the experimental observations from two distinct incommensurate CDW systems, we conjecture that strong el-ph coupling, including a strong dependence on the electronic momentum as well as the orbital character of the participating electronic states are the defining attributes of  the incommensurate CDW orders in TMDs. 

~\\ \indent
{\bf Acknowledgments} \\
U.C. acknowledges supports from the National Science Foundation under Grant No. DMR-1454304 and from the Jefferson Trust at the University of Virginia. Work at Argonne National Laboratory (C.D.M., D.Y.C., M.G.K.) was supported by the U.S. Department of Energy, Office of Basic Energy Sciences, Division of Materials Science and Engineering. D.L is supported by the Department of Energy, Grant number DE-FG02-01ER45927. J.v.W. acknowledges support from a VIDI grant financed by the Netherlands Organisation for Scientific Research (NWO).

\end{document}